\documentclass[runningheads]{llncs}
\usepackage[T1]{fontenc}
\usepackage{microtype}
\usepackage{graphicx}
\usepackage[colorlinks=true, citecolor=blue, linkcolor=blue, urlcolor=blue]{hyperref}
\usepackage{amsmath}
\usepackage{algorithm}
\usepackage{algorithmic}
\usepackage{amsfonts}
\usepackage{adjustbox}
\usepackage{hyperref}
\usepackage{marvosym}

\begin{document}
\title{BRIDG-Q: Barren-Plateau-Resilient Initialisation with Data-Aware LLM-Generated Quantum Circuits}
\titlerunning{BRIDG-Q: A Quantum Circuit Initialisation Framework}
%
\author{
Ngoc Nhi Nguyen\orcidID{0009-0006-8228-3615}(\Letter) \and
Thai T. Vu\orcidID{0000-0002-9826-3321} \and
John Le\orcidID{0000-0003-0019-0345} \and
Hoa Khanh Dam\orcidID{0000-0003-4246-0526} \and
Dung Hoang Duong\orcidID{0000-0001-8057-4060} \and
Dinh Thai Hoang\orcidID{0000-0002-9528-0863}
}

\authorrunning{N. N. Nguyen et al.}

\institute{
University of Wollongong, Wollongong, NSW, Australia\\
\email{nnn469@uowmail.edu.au}, 
\email{\{tienv,johnle,hoa,hduong\}@uow.edu.au}
\and
University of Technology Sydney, Sydney, NSW, Australia\\
\email{Hoang.Dinh@uts.edu.au}
}
\maketitle              
\begin{abstract}
Quantum circuit initialisation is a key bottleneck in variational quantum algorithms (VQAs), strongly impacting optimisation stability and convergence. Recent work shows that large language models (LLMs) can synthesise high-quality variational circuit architectures, but their continuous parameter predictions are unreliable. Conversely, data-driven initialisation methods such as BEINIT improve trainability via problem-adaptive priors, yet assume fixed ansatz templates and ignore generative circuit structure. We propose \textbf{BRIDG-Q} (\textbf{B}arren-Plateau-\textbf{R}esilient \textbf{I}nitialisation with \textbf{D}ata-Aware LLM-\textbf{G}enerated \textbf{Q}uantum Circuits), a neuro-symbolic pipeline that bridges this gap by coupling LLM-generated circuit architectures with empirical-Bayes parameter initialisation. BRIDG-Q uses AgentQ to generate problem-conditioned circuit topologies, removes generated parameters, and injects data-informed parameter initialisations to mitigate barren plateau effects. Evaluations on graph optimisation benchmarks using residual energy gap and convergence metrics show improved optimisation robustness, indicating that data-driven initialisation remains effective even for LLM-generated circuits, with oracle per-instance selection achieving approximately a $10\%$ reduction in final residual energy.

\paragraph{Reproducibility.}
All source code and data used in this study
are publicly available at:
\href{https://github.com/nellyy2505/BRIDG-Q}{\texttt{https://github.com/nellyy2505/BRIDG-Q}}
\end{abstract}

\keywords{Variational quantum algorithms; barren plateaus; parameter initialisation; empirical Bayes; Beta distribution; large language models; quantum circuit generation; neuro-symbolic optimisation.}

\section{Introduction}

Variational Quantum Algorithms (VQAs) are a leading approach for near-term quantum computation, enabling hybrid quantum--classical optimisation on noisy intermediate-scale quantum (NISQ) devices \cite{Preskill2018NISQ,Moll_2018}. In a typical VQA, a parameterised quantum circuit (PQC) prepares a quantum state whose parameters are iteratively updated by a classical optimiser to minimise a problem-dependent cost function, such as the expectation value of a Hamiltonian \cite{Cerezo2021VQA,PhysRevA.98.032309}. Despite their expressive power, VQAs are highly sensitive to circuit design choices, optimisation dynamics, and, critically, the initialisation of variational parameters \cite{Qi2024VQAReview}.

A central obstacle to scalable VQA training is the \emph{barren plateau} phenomenon, in which the variance of cost-function gradients decays exponentially with system size, rendering gradient-based optimisation ineffective \cite{McClean2018Barren}. Subsequent studies have shown that barren plateaus are not only determined by circuit depth, but also strongly influenced by entanglement structure and the locality of the cost function \cite{Cerezo2021Cost,Marrero2021EntanglementBP}. As a result, uninformed parameter initialisation can cause optimisation to stagnate early, even when the circuit itself is sufficiently expressive.

Two largely independent research directions have recently emerged to address complementary aspects of this challenge. On the one hand, data-driven initialisation strategies construct problem-adaptive priors over circuit parameters, significantly improving gradient behaviour during training \cite{Kulshrestha2022BEINIT,Rad2022BayesianInit}. On the other hand, large language models (LLMs) have been shown to generate expressive variational circuit architectures directly from problem descriptions \cite{Jern2025AgentQ,10691762}. However, existing pipelines typically combine LLM-generated circuit structures with generic or heuristic parameter initialisation, while data-driven initialisation methods assume fixed ansatz templates and do not exploit generative circuit synthesis. As a result, circuit architecture generation and parameter initialisation remain largely decoupled in end-to-end VQA workflows.

In this work, we bridge these two lines of research through a \emph{neuro-symbolic} approach \cite{10.1093/nsr/nwac035}. In our setting, ``neuro-symbolic'' denotes the combination of \textbf{neural reasoning}, embodied by an LLM that synthesises variational circuit topologies from data and examples, and \textbf{symbolic inductive bias}, encoded through human-interpretable rules for parameter initialisation grounded in empirical Bayes principles. This design complements the flexibility of learned structure generation with principled control over optimisation dynamics.

Our framework provides the following contributions:
\begin{enumerate}
    \item We introduce \textbf{BRIDG-Q}, a unified pipeline that couples
    LLM-generated variational circuit architectures with empirical-Bayes parameter initialisation. This approach explicitly addresses the disconnect between
    circuit synthesis and parameter optimisation in end-to-end VQA workflows.
    \item Extending BEINIT \cite{Kulshrestha2022BEINIT}, we propose a \emph{gate-aware stratified Beta} initialisation scheme that applies role-specific parameter scaling to
    control early entanglement growth, hence mitigating barren plateau effects in deep circuits.
    \item We conduct a paired empirical evaluation on approximately
    $580$ graph-based optimisation instances from the AgentQ benchmark suite \cite{Jern2025AgentQ},
    using energy-gap and convergence-efficiency metrics to characterise both
    solution quality and optimisation dynamics.
\end{enumerate}

The structure of this paper is as follows. First, we review background on barren plateaus in variational quantum algorithms, data-driven parameter initialisation, and recent progress in LLM-based circuit generation. We then introduce the BRIDG-Q framework, describing how LLM-generated circuit structures are combined with empirical-Bayes parameter initialisation, including the gate-aware stratified variant. Next, we define the evaluation metrics and experimental protocol used to assess both solution quality and convergence behaviour. We subsequently present paired empirical results on graph-based optimisation instances from the AgentQ benchmark, comparing BRIDG-Q variants against standard baselines. Finally, we discuss the implications of these findings, outline current limitations, and highlight directions for future work.

\vspace{0.6em}
\section{Related Work}
\paragraph{Barren plateaus and parameter initialisation.}
The scalability of variational quantum algorithms (VQAs) is fundamentally limited by optimisation pathologies such as barren plateaus, first formally identified by McClean \emph{et al.}, who showed that gradient variance can decay exponentially with the number of qubits under random parameter initialisation \cite{McClean2018Barren}. Subsequent analyses by Cerezo \emph{et al.} demonstrated that barren plateau behaviour depends on the locality of the cost function and the circuit structure \cite{Cerezo2021Cost}, while Marrero \emph{et al.} linked plateau formation directly to entanglement growth and the emergence of unitary 2-designs \cite{Marrero2021EntanglementBP,PhysRevB.101.144305}. These results motivated a line of work aimed at controlling circuit expressivity, particularly during the early stages of training. A number of heuristic and geometric initialisation strategies have been proposed to mitigate these effects \cite{garciasaez2018addressinghardclassicalproblems}. Grant \emph{et al.} introduced \emph{identity initialisation}, in which parameters are chosen so that the circuit evaluates close to the identity operator, thereby preserving gradient magnitudes at initialisation \cite{Grant2019Initialisation}. More recently, Zhang \emph{et al.} proposed \emph{Gaussian initialisation} schemes with depth-dependent variance, providing theoretical guarantees that gradients vanish at most polynomially with circuit depth \cite{Zhang2022Gaussian}. While these methods offer important insights into trainability, they do not exploit problem-specific structure encoded in the input instance.

\paragraph{Empirical-Bayes initialisation.}
BEINIT \cite{Kulshrestha2022BEINIT} proposes an empirical-Bayes approach to parameter initialisation, in which circuit parameters are sampled from a Beta distribution whose hyperparameters are estimated from problem data via maximum likelihood. This strategy demonstrably increases gradient variance and reduces the likelihood of barren plateaus in deep variational circuits. However, BEINIT assumes a fixed ansatz structure and treats circuit parameters homogeneously, without accounting for gate semantics or circuit topology generated dynamically. As a result, its applicability is limited to scenarios in which the circuit architecture is predefined.

\paragraph{LLM-based circuit generation.}
Recent work has shown that LLMs can generate expressive variational circuit topologies directly from problem descriptions, reducing the need for manual ansatz design \cite{PhysRevA.98.032309,Ostaszewski2021structure}. AgentQ \cite{Jern2025AgentQ} exemplifies this direction by synthesising problem-conditioned circuits that achieve competitive performance across a range of graph-based optimisation tasks. Nevertheless, existing pipelines generate circuit structure and continuous parameters jointly, without explicitly modelling parameter initialisation as a separate problem-adaptive design stage. As a result, the generated parameters do not incorporate fitted problem-aware priors or trainability-driven initialisation principles.

\section{Methodology: The BRIDG-Q Framework}

We propose \textbf{BRIDG-Q}, a neuro-symbolic framework that unifies generative topology design with data-driven parameter initialisation. The architecture addresses the ``compatibility gap'' between the discrete, symbolic reasoning of LLMs and the continuous, geometric optimisation required for VQAs.

\begin{figure}[htbp]
  \centering
  \includegraphics[width=0.95\linewidth]{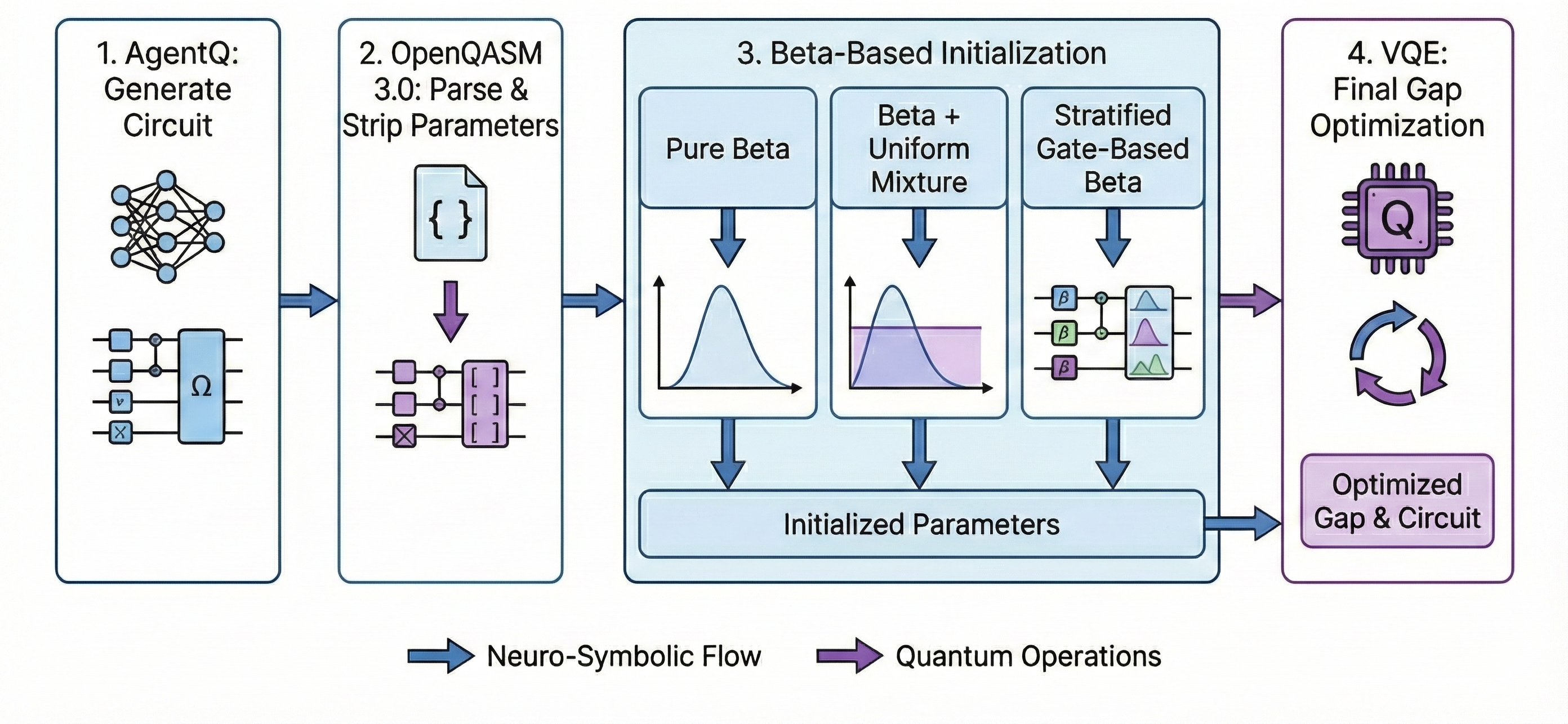}
  \caption{Overview of the BRIDG-Q neuro-symbolic pipeline for variational quantum optimisation. Figure created with assistance from Gemini~3 and edited by the authors.}
  \label{fig:initq-architecture}
\end{figure}

\vspace{-2.5em}
\subsection{Architectural Overview}

Figure~\ref{fig:initq-architecture} illustrates the BRIDG-Q pipeline as a
directed acyclic workflow that transforms a graph-based combinatorial
optimisation problem into an optimised variational quantum circuit. The design
explicitly separates \emph{circuit structure generation}, \emph{parameter
initialisation}, and \emph{variational optimisation}. The workflow proceeds
through four stages:

\begin{enumerate}
    \item \textbf{LLM-Based Circuit Generation (AgentQ).}  
    In the first stage, an LLM, AgentQ, is used to generate the
    \emph{structure} of the variational ansatz. Conditioned on the input problem
    description, AgentQ produces an
    OpenQASM~3.0 \cite{10.1145/3505636} circuit specifying the gate sequence, qubit connectivity, and
    overall circuit layout. This stage is purely structural: although AgentQ emits numerical parameters, these are not retained, as they are not optimised with respect to the variational objective. 

    \item \textbf{OpenQASM Parsing and Parameter Stripping.}  
    The generated OpenQASM~3.0 \cite{10.1145/3505636} code is then parsed to recover the circuit
    topology and identify all parameterised gates. Any numerical parameter values produced by the LLM are removed at this stage. The output of this step is a parameter-free circuit template that preserves the LLM-generated structure while deferring all continuous parameter
    choices to a dedicated initialisation module.

    \item \textbf{Beta-Based Parameter Initialisation.}  
    In the third stage, BRIDG-Q injects parameters into the stripped circuit using data-driven Beta priors. Problem-specific features, such as graph edge
    weights and Hamiltonian coefficients, are extracted from the input instance and normalised. These features are used to fit a Beta distribution via
    maximum likelihood estimation, following the BEINIT principle. From this learned prior, parameters are sampled using one of three variants:
    \emph{pure Beta sampling}, a \emph{Beta--uniform mixture}, or a
    \emph{gate-aware stratified Beta scheme}.

    \item \textbf{Variational Optimisation (VQE).}  
    The initialised circuit is optimised using a standard VQE loop \cite{Peruzzo2014VQE}, yielding optimised parameters and a final residual energy gap.
\end{enumerate}

\paragraph{Illustrative example: Max-Cut.}
Consider a four-node weighted Max-Cut instance with edges
$\{(0,1,\,0.7),(1,2,\,0.3),(2,3,\,0.9),(0,3,\,0.5)\}$ and Hamiltonian
$H=\sum_{(i,j)}w_{ij}\tfrac{1}{2}(I-Z_iZ_j)$. In Stage~1, AgentQ
produces a circuit with four \texttt{ry} and three \texttt{crz} gates. In
Stage~2, all numerical angles are stripped, leaving seven empty parameter
slots. In Stage~3, the normalised edge weights
$[0.7,0.3,0.9,0.5]$ yield a fitted prior
$\text{Beta}(1.8,1.4)$; under the stratified variant, \texttt{ry}
parameters are mapped to $[-\pi,\pi]$ while \texttt{crz} parameters are
confined to $[-0.2,0.2]$. Stage~4 then optimises via a standard VQE loop.

\subsection{Data-Driven Priors: The BEINIT Formulation}

As outlined in the architectural overview, BRIDG-Q explicitly removes all
LLM-generated numerical parameters, leaving a symbolic circuit template.
Section~3.2 describes how this template is populated with data-driven initial
parameters using an empirical-Bayes formulation based on BEINIT \cite{Kulshrestha2022BEINIT}.

\paragraph{Feature extraction.}
The key idea behind this empirical Bayes formulation is to encode structural information about the optimisation problem into the parameter prior. Concretely, we construct a feature vector $\mathbf{x}_{\text{feat}}$ by aggregating information from two complementary sources:
\begin{itemize}
    \item the \textbf{graph structure} of the problem instance, such as edge weights or capacities in the combinatorial graph, and
    \item the \textbf{Hamiltonian coefficients} extracted from the problem’s cost function, provided by AgentQ as a \texttt{cost\_hamiltonian} string.
\end{itemize}
All extracted values are parsed, normalised to the interval $[0,1]$, and concatenated into a single empirical data vector. Although BEINIT originally estimates its prior from labelled classical datasets, we show that graph- and Hamiltonian-derived features serve as an effective surrogate, enabling instance-specific prior construction without supervised labels.

\paragraph{Beta distribution via MLE.}
Given the normalised feature vector $\{x_i\}_{i=1}^N$, we fit a Beta distribution
using maximum likelihood estimation (MLE), where each $x_i \in [0,1]$ denotes a
scalar feature extracted from the input instance. The Beta density is defined as
\begin{equation}
    f(x; \alpha, \beta) \;=\; \frac{x^{\alpha-1}(1-x)^{\beta-1}}{B(\alpha,\beta)},
\end{equation}
where $\alpha$ and $\beta$ are shape parameters and $B(\alpha,\beta)$ denotes the Beta function. The optimal parameters are obtained by solving
\begin{equation}
    (\hat{\alpha}, \hat{\beta}) 
    \;=\;
    \underset{\alpha,\beta}{\arg\max}
    \sum_{i=1}^{N} \log f(x_i; \alpha, \beta).
\end{equation}
In practice, we perform a constrained Beta fit with fixed location and scale, and introduce a small numerical floor to avoid degenerate solutions. If the fitting procedure fails or the extracted features are constant, the method falls back to $\hat{\alpha} = \hat{\beta} = 1$, which corresponds to a uniform distribution.

\paragraph{Initialisation variants.}
Building on the learned Beta prior, we implement the following \emph{initialisation-only} variants within BRIDG-Q:
\begin{itemize}
    \item \textbf{$\text{BRIDG-Q}^{\beta\text{--pure}}$.}  
    All parameters $\theta_k$ are sampled directly from $\text{Beta}(\hat{\alpha}, \hat{\beta})$, then mapped to gate-specific angular ranges. This injects a strong inductive bias derived from the problem instance while retaining full expressive capacity.

    \item \textbf{$\text{BRIDG-Q}^{\beta\text{--mixture}}$.}  
    To guard against overconfident or collapsed priors, we mix the learned Beta distribution with a uniform component:
    \begin{equation}
        \theta_k \sim (1-\lambda)\,\text{Beta}(\hat{\alpha}, \hat{\beta}) 
        \;+\;
        \lambda\,\mathcal{U}[0,1],
    \end{equation}
    where $\lambda = 0.2$. This introduces controlled stochasticity at initialisation. While BEINIT injects Gaussian noise during training to escape saddle points \cite{Kulshrestha2022BEINIT}, our use of a uniform mixture is restricted strictly to initialisation and serves to preserve exploration without modifying the optimisation dynamics.
\end{itemize}

\subsection{$\text{BRIDG-Q}^{\beta\text{--stratified}}$: Gate-Aware Initialisation}

While standard Beta-based initialisation already improves trainability by shaping the initial parameter distribution according to the problem instance, it still applies a uniform scaling rule across all parameterised gates. In deep circuits, this can lead to excessive entanglement at initialisation, accelerating the formation of unitary 2-designs and inducing barren plateaus \cite{Patti2021EntanglementDevised}. To address this limitation, we introduce \textbf{$\text{BRIDG-Q}^{\beta\text{--stratified}}$} which follows this principle: it restricts the initial strength of entangling gates while allowing single-qubit \emph{driver} gates to fully explore the parameter space.

\paragraph{Core idea.}
The key idea is to decouple where parameters come from (the Beta distribution learned via BEINIT) from how they are applied (gate-specific scaling rules). All parameters are sampled from the same Beta prior, then mapped to gate-dependent angle ranges based on functional role.

\begin{algorithm}[t]
\caption{BRIDG-Q Initialisation}
\label{alg:bridgq_init}
\begin{algorithmic}[1]
\REQUIRE Circuit description $Q$; problem instance $(G,H)$
\ENSURE Initial parameter vector $\boldsymbol{\theta}_{\text{init}}$
\STATE $C \gets \textsc{ParseOpenQASM}(Q)$ \COMMENT{Recover gate list, wires, and param slots}
\STATE $C^{\emptyset} \gets \textsc{StripParameters}(C)$ \COMMENT{Discard all LLM-emitted numeric values}
\STATE $\mathbf{x}_{\text{feat}} \gets \textsc{ExtractFeatures}(G,H)$
\STATE Normalise $\mathbf{x}_{\text{feat}}$ to $[0,1]$
\STATE $(\hat{\alpha},\hat{\beta}) \gets \textsc{MLE\_BetaFit}(\mathbf{x}_{\text{feat}})$
\STATE $\boldsymbol{v} \gets \textsc{SampleBeta}(\hat{\alpha},\hat{\beta},\,|\mathcal{P}|)$ \COMMENT{$|\mathcal{P}|$ = \# parameterised gates}
\IF{variant = $\beta$--mixture}
    \STATE Replace a fraction $\lambda$ of entries in $\boldsymbol{v}$ with $\mathcal{U}[0,1]$
\ENDIF
\STATE $\boldsymbol{\theta}_{\text{init}} \gets [\,]$; $p \gets 1$
\FOR{each gate $g \in C^{\emptyset}$ in program order}
    \IF{$g$ is parameterised}
        \STATE $v \gets \boldsymbol{v}[p]$; $p \gets p+1$
        \IF{variant = $\beta$--stratified}
            \IF{$g.\text{type} \in \{\texttt{rx},\texttt{ry},\texttt{u3}\}$} 
                \STATE $\theta \gets 2\pi v - \pi$
            \ELSE 
                \STATE $\theta \gets (v-0.5)\cdot \epsilon$
            \ENDIF
        \ENDIF
        \STATE Append $\theta$ to $\boldsymbol{\theta}_{\text{init}}$
    \ENDIF
\ENDFOR
\RETURN $\boldsymbol{\theta}_{\text{init}}$
\end{algorithmic}
\end{algorithm}
\paragraph{Algorithmic procedure.}
Algorithm~\ref{alg:bridgq_init} summarises the BRIDG-Q initialisation process.
Problem-specific features are extracted from the graph $G$ and Hamiltonian $H$,
normalised to $[0,1]$, and used to fit a Beta distribution via maximum likelihood
estimation, yielding shape parameters $(\hat{\alpha},\hat{\beta})$
(lines~3--5). The circuit is then traversed in program order
(lines~11--23). For each parameterised gate, a latent value is sampled from the
fitted Beta prior (line~6), the functional role of the gate is identified, and a
role-specific scaling transformation is applied to obtain the initial parameter
value $\theta_k$ (lines~12--19). The resulting parameters are assembled into the
initial vector $\boldsymbol{\theta}_{\mathrm{init}}$ returned for optimisation
(lines~21--24).

\paragraph{Gate-specific scaling rules.}
In practice, the stratification is implemented as follows:
\begin{itemize}
    \item \textbf{Driver gates} (e.g., \texttt{rx}, \texttt{ry}, \texttt{u3}) receive
    parameters mapped to the full angular range,
    \begin{equation}
        \theta_k = 2\pi v_k - \pi ,
        \label{eq:driver-scaling}
    \end{equation}
    corresponding to $[-\pi, \pi]$. This allows these gates to immediately explore
    the full Bloch sphere and drive meaningful state evolution.

    \item \textbf{Entangling gates} (e.g., \texttt{crz}, \texttt{crx}, \texttt{cry})
    are initialised near the identity,
    \begin{equation}
        \theta_k = (v_k - 0.5)\,\epsilon ,
        \label{eq:entangler-scaling}
    \end{equation}
    where $\epsilon = 0.4$ by default, yielding a narrow range of $[-0.2, 0.2]$.
    This ensures that early entanglement is weak and gradually increases.
\end{itemize}

\paragraph{Rationale.}
By initialising entangling operations near the identity, $\text{BRIDG-Q}^{\beta\text{--stratified}}$ induces a weakly entangled starting regime even for deep circuits. This delays early randomisation and mitigates gradient collapse associated with approximate unitary 2-designs \cite{Patti2021EntanglementDevised}. As optimisation proceeds, entanglement increases only where supported by gradient descent, enabling a smooth transition from a highly trainable regime to a more expressive circuit.

\paragraph{Remarks.}
Some single-qubit gates (e.g., \texttt{rz}) are conservatively scaled to prioritise stability at initialisation. This choice slightly restricts expressivity at $t=0$ but does not limit the reachable parameter space during training.

\section{Evaluation Metrics}
We evaluate (i) solution quality via the residual energy gap and (ii) optimisation efficiency via iterations- and time-to-convergence. In practice, we report the evolution of the energy gap across optimisation steps and define it as
\begin{equation}
\Delta E(\boldsymbol{\theta}_t) \;=\; \left| E(\boldsymbol{\theta}_t) - E_{\mathrm{exact}} \right|,
\label{eq:energy-gap}
\end{equation}
where $E(\boldsymbol{\theta}_t)$ denotes the variational objective value at optimisation step $t$. $E_{\mathrm{exact}}$ is the ground-truth optimal energy for the corresponding problem instance \cite{Anschuetz2022Traps}.

\subsection{Primary metric: Energy gap}

Here the variational objective is defined as
\begin{equation}
E(\boldsymbol{\theta})
\;=\;
\langle \psi(\boldsymbol{\theta}) \lvert H \rvert \psi(\boldsymbol{\theta}) \rangle,
\end{equation}
where $H$ denotes the problem Hamiltonian. Throughout this work, we use the absolute difference in Eq.~\eqref{eq:energy-gap}, which directly corresponds to the curves reported in our experimental results.

The energy gap provides a direct and task-aligned measure of performance. Rather than focusing on optimiser-specific signals, it evaluates whether the learned parameters actually produce a low-energy state for the target Hamiltonian \cite{Peruzzo2014VQE}. This distinction is particularly important in the presence of barren plateaus: an optimiser may appear to converge or stabilise while remaining far from the true optimum \cite{Cerezo2021Cost,Marrero2021EntanglementBP}. By tracking $\Delta E(\boldsymbol{\theta}_t)$ over time, we capture both 
(i) \emph{final solution quality}, where a smaller terminal gap indicates a better approximation to the optimal solution, and  
(ii) \emph{convergence behavior}, reflected in how rapidly the energy gap decreases during optimisation.

The reference value $E_{\mathrm{exact}}$ is taken from the AgentQ benchmark suite, which provides a classically computed optimum objective value for each instance \cite{Jern2025AgentQ}. This fixed reference enables paired, instance-wise comparisons across initialisation strategies under identical problem instances.

\subsection{Secondary metrics: Optimisation efficiency}
Energy-gap curves convey both quality and dynamics, but two additional metrics are useful for quantifying efficiency in a single scalar per run.

\subsubsection{Iteration steps.}
We record the number of optimisation iterations executed, and in particular the \emph{iterations-to-converge}:
\begin{equation}
T_{\mathrm{conv}} \;=\; \min\left\{t \;:\; \Delta E(\boldsymbol{\theta}_t) \le \varepsilon \right\},
\label{eq:iter-to-converge}
\end{equation}
for a chosen tolerance $\varepsilon$. If the run never satisfies the criterion within a maximum budget $T_{\max}$, we set $T_{\mathrm{conv}} = T_{\max}$ and treat it as non-convergent under the given budget.
Since each iteration entails multiple circuit evaluations, $T_{\mathrm{conv}}$ serves as a proxy for total quantum-resource usage.

\subsubsection{Time to converge.}
We measure wall-clock \emph{time-to-converge} as
\begin{equation}
t_{\mathrm{conv}} \;=\; \text{elapsed wall-clock time until Eq.~\eqref{eq:iter-to-converge} is first satisfied},
\end{equation}
including all forward evaluations (expectation estimation), gradient evaluations (if applicable), and classical optimiser overhead. This captures implementation-dependent costs not reflected by iteration counts.

\section{Experimental Results}

We evaluate BRIDG-Q using a paired comparison protocol in which all methods are tested on identical problem instances and circuit topologies, differing only in parameter initialisation.

\subsection{Experimental Setup}
\label{sec:exp_setup}

\paragraph{Benchmark.}
Experiments are conducted on approximately $580$ graph-based optimisation instances from the AgentQ benchmark \cite{Jern2025AgentQ}, each consisting of a fixed problem Hamiltonian and an AgentQ-generated variational circuit expressed in OpenQASM~3.0. Following the AgentQ evaluation protocol, instances with missing baseline results or final baseline energy gap exceeding $5.0$ are excluded, leaving $551$ paired instances for analysis.

\paragraph{Initialisation strategies.}
We compare the following methods, all evaluated on identical circuit topologies:
\begin{itemize}
    \item \textbf{AgentQ (baseline):} LLM-generated parameters used directly.
    \item \textbf{Random:} Independent uniform sampling.
    \item \textbf{Uniform:} Deterministic uniform grid initialisation.
    \item $\textbf{BRIDG-Q}^{\beta\text{--pure}}$: Sampling from a fitted Beta prior.
    \item $\textbf{BRIDG-Q}^{\beta\text{--mixture}}$: Beta--uniform mixture ($\lambda=0.2$).
    \item $\textbf{BRIDG-Q}^{\beta\text{--stratified}}$: Gate-aware Beta initialisation.
    \item $\textbf{BRIDG-Q}^{\beta\text{--best}}$: Oracle per-instance selection among Beta variants. This variant is reported solely to estimate
    the upper bound of Beta-based initialisation performance, and is not a
    deployable method.
\end{itemize}

\paragraph{Optimisation settings.}
All methods share identical optimisation hyperparameters, summarised in Table~\ref{tab:exp_config}.

\begin{table}[htbp]
\centering
\caption{Experimental configuration shared across all evaluated initialisation strategies.}
\small
\setlength{\tabcolsep}{6pt}
\renewcommand{\arraystretch}{1.1}
\begin{tabular}{ll}
\hline
\textbf{Parameter} & \textbf{Value} \\
\hline
Optimiser & Adam \cite{Kingma2014Adam} \\
Maximum iterations ($T_{\max}$) & 400 \\
Convergence threshold ($\varepsilon$) & $0.05$ (energy gap) \\
Baseline initialisation & AgentQ-generated \\
Beta mixture coefficient ($\lambda$) & $0.2$ \\
Entangler scaling ($\epsilon$) & $0.4$ \\
Beta fitting method & MLE (fixed support) \\
Evaluation protocol & Paired, per-instance \\
\hline
\end{tabular}
\label{tab:exp_config}
\end{table}

\paragraph{Implementation and compute environment.}
All experiments are implemented in Python and executed on a Google Cloud Platform VM (\texttt{n1-standard-8}: 8~vCPUs, 30\,GB RAM) equipped with a single NVIDIA T4 GPU, using Python~3.10 and CUDA~12.4.

\subsection{Evaluation Results}
Table~\ref{tab:paired_results} reports the paired performance of all
initialisation strategies across $551$ problem instances after baseline
filtering. Results are summarised using mean residual energy, median paired
improvement relative to the AgentQ baseline, success probability, and convergence
latency, and wall-clock time.

\begin{table}[htbp]
\centering
\caption{Paired comparison of initialisation strategies under identical problem instances.}
\small
\setlength{\tabcolsep}{3.5pt}
\renewcommand{\arraystretch}{1.05}
\begin{adjustbox}{max width=\linewidth,center}
\begin{tabular}{lccccc}
\hline
Initialisation Method &
Mean Residual &
Median Improv. &
Success Prob. &
Conv. Latency &
Time (s) \\
\hline
AgentQ (baseline) &
$0.440 \pm 0.853$ &
--- &
$65.0\%$ &
$236.3$ &
$1.99$ \\

Random &
$1.453 \pm 4.079$ &
$-43.14\%$ &
$42.6\%$ &
$296.2$ &
$2.70$ \\

Uniform &
$1.447 \pm 3.787$ &
$-37.86\%$ &
$44.6\%$ &
$314.5$ &
$2.67$ \\

$\text{BRIDG-Q}^{\beta\text{--pure}}$ &
$1.203 \pm 2.967$ &
$-18.23\%$ &
$45.9\%$ &
$331.7$ &
$2.66$ \\

$\text{BRIDG-Q}^{\beta\text{--mixture}}$ &
$1.147 \pm 3.265$ &
$-20.61\%$ &
$44.8\%$ &
$323.6$ &
$2.72$ \\

$\text{BRIDG-Q}^{\beta\text{--stratified}}$ &
$1.373 \pm 4.150$ &
$-7.39\%$ &
$49.9\%$ &
$342.9$ &
$2.73$ \\

$\text{BRIDG-Q}^{\beta\text{--best}}$ &
$\mathbf{0.368 \pm 1.522}$ &
$\mathbf{+10.39\%}$ &
$\mathbf{69.7\%}$ &
$338.6$ &
$2.79$ \\
\hline
\end{tabular}
\end{adjustbox}
\label{tab:paired_results}
\end{table}

\paragraph{Paired evaluation protocol and filtering.}
A run is considered successful if the final residual energy gap satisfies
$\mathrm{gap}_{\mathrm{final}} \le 0.05$.
Instances for which the baseline initialisation produces a final gap greater
than $5.0$ (or no valid result) are excluded.
All reported statistics are computed using strict per-instance pairing: a method
contributes only when both it and the baseline produce valid energy gap and
runtime measurements.
We additionally report an oracle-selected variant,
$\textbf{BRIDG-Q}^{\beta\text{--best}}$, which retrospectively chooses the
best-performing BRIDG-Q variant per instance based on final gap
(ties broken by runtime).

\paragraph{Baseline and uninformed strategies.}
AgentQ-generated initialisation provides a strong baseline, achieving a mean
residual energy of $0.440$ with a $65.0\%$ success probability. In contrast,
uninformed strategies such as random and uniform initialisation perform
substantially worse across all metrics. Both exhibit large residual energies,
negative median paired improvements (below $-35\%$), and markedly lower success
rates, confirming that naive initialisation is insufficient. We note that the large standard deviations in Table~\ref{tab:paired_results} reflect high instance-level variability inherent to the diverse graph structures in the benchmark; however, the consistent directionality of median paired improvements across all $551$ instances, combined with the substantial gaps in success probability, suggest that the observed differences are unlikely to be attributable to noise alone.

\paragraph{Individual Beta-based strategies.}
When evaluated individually, Beta-based initialisations (\texttt{$\text{BRIDG-Q}^{\beta\text{--pure}}$},
\texttt{$\text{BRIDG-Q}^{\beta\text{--mixture}}$}, and \texttt{$\text{BRIDG-Q}^{\beta\text{--stratified}}$}) improve upon random and
uniform baselines but do not consistently outperform AgentQ. As shown in
Table~\ref{tab:paired_results}, their median paired improvements remain negative,
indicating that no single fixed Beta prior is optimal across the full dataset.

\paragraph{Oracle-selected Beta initialisation.}
To assess the upper bound of Beta-based initialisation performance, we report an
oracle-selected variant, \textbf{$\text{BRIDG-Q}^{\beta\text{--best}}$}. For each instance,
the oracle retrospectively selects the best-performing Beta-based strategy based
on the final energy gap (with runtime as a tie-breaker). While not deployable, this
analysis reveals that for a substantial fraction of instances, at least one
Beta-based initialisation outperforms the AgentQ baseline.

Quantitatively, the oracle-selected Beta strategy achieves the highest success
probability ($69.7\%$) and a positive median paired improvement of
$+10.39\%$, representing an approximate $10\%$ reduction in residual energy
relative to AgentQ (Table~\ref{tab:paired_results}). These results indicate that the limitation of Beta-based
initialisation lies not in the expressiveness of the priors themselves, but in the
absence of an instance-aware mechanism for selecting the most suitable prior.

\begin{figure}[htbp]
    \centering
    \includegraphics[width=0.6\linewidth]{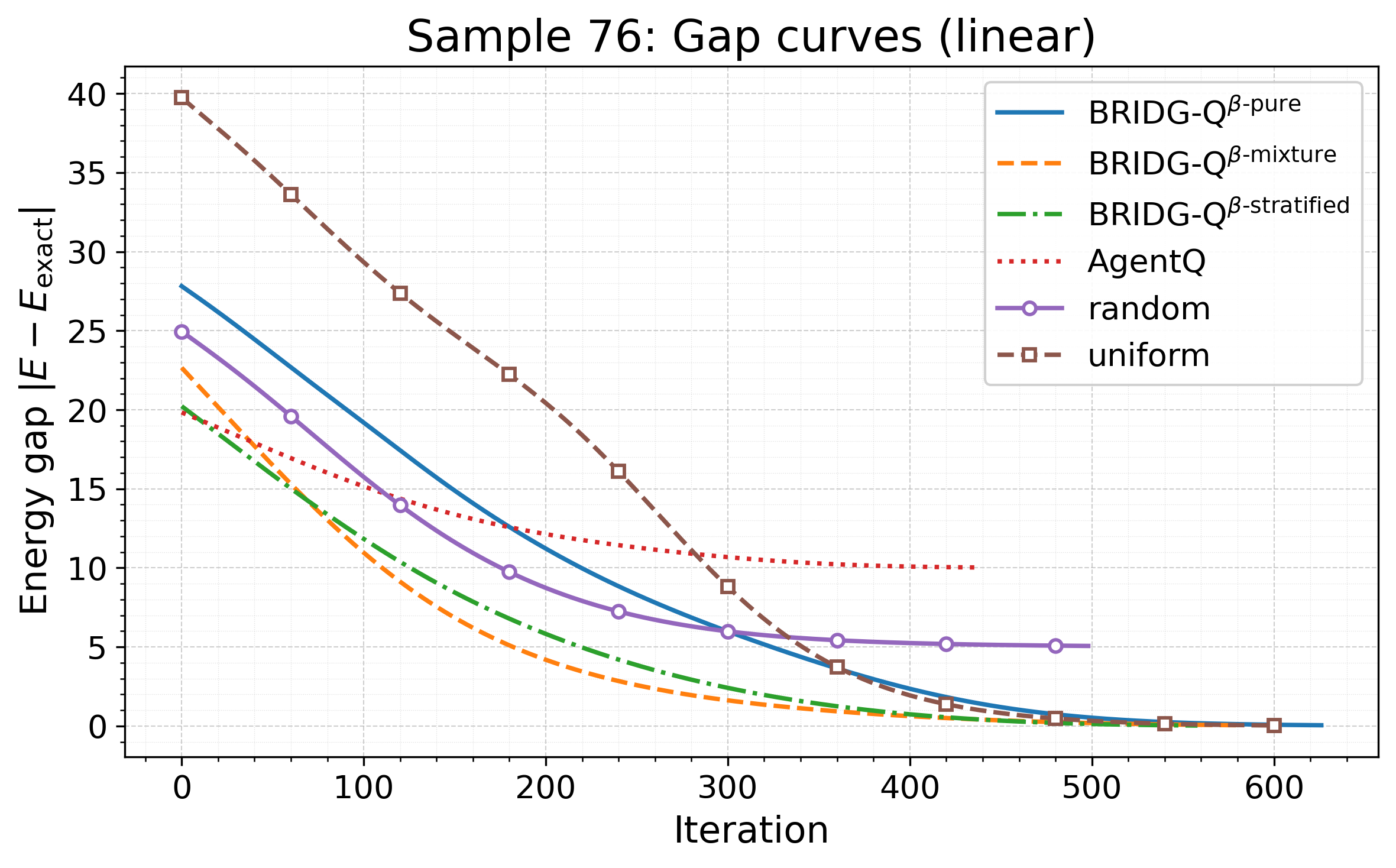}
    \caption{Energy-gap trajectories for a representative problem instance (Sample 76). The plot shows the evolution of the absolute energy gap over optimisation iterations for different initialisation strategies. Beta-based methods ($\text{BRIDG-Q}^{\beta\text{--pure}}$, $\text{BRIDG-Q}^{\beta\text{--mixture}}$, and $\text{BRIDG-Q}^{\beta\text{--stratified}}$) exhibit slower initial convergence but continue to reduce the energy gap beyond the point where standard baselines (AgentQ, random, uniform) stagnate. In particular, $\text{BRIDG-Q}^{\beta\text{--stratified}}$ achieves the lowest final gap for this instance}
    
    \label{fig:samplecurve}
\end{figure}

\subsection{Discussion}

\paragraph{Why Beta-based methods lose individually.}
The inconsistent performance can be
attributed to instance heterogeneity. Different graph structures and Hamiltonian
distributions induce distinct optimisation landscapes \cite{krentel1986complexity}, for which a single prior shape may be mismatched. As a result, fixed Beta priors may either under-regularise or over-constrain certain instances, leading to suboptimal convergence despite improvements over uninformed initialisation.

\paragraph{What the oracle result reveals.}
The strong performance of the oracle-selected Beta variant demonstrates that
data-driven priors are frequently beneficial, but their effectiveness is
instance-dependent. The oracle exposes latent performance potential by showing
that, for many instances, a suitable Beta-based initialisation exists that
surpasses the LLM-generated baseline. This suggests that future gains depend less
on inventing new priors and more on learning how to select or adapt priors based
on instance-level signals.

\paragraph{Convergence behaviour.}
Beta-based methods typically require more optimisation steps than the AgentQ baseline, reflecting prolonged exploration rather than inefficiency. By delaying early entanglement and gradient collapse, these methods maintain informative gradients for longer, which increases the likelihood of reaching lower-energy solutions.

\paragraph{Practical considerations.}
Despite longer optimisation trajectories, wall-clock runtimes remain comparable across methods, indicating that improved robustness does not incur substantial computational overhead.

\section{Conclusion}
We presented \textbf{BRIDG-Q}, a neuro-symbolic framework that decouples LLM-based variational circuit synthesis from parameter initialisation in end-to-end VQA workflows. By injecting empirical-Bayes, gate-aware parameter priors into automatically generated circuit structures, BRIDG-Q addresses the mismatch between discrete circuit generation and continuous optimisation. Paired evaluation on a large AgentQ benchmark shows that while no single Beta-based strategy consistently outperforms the AgentQ baseline, oracle per-instance selection achieves an approximately $10\%$ reduction in final residual energy with comparable runtime, indicating improved robustness in deep VQAs. These results highlight both the effectiveness and limitations of fixed, hand-designed priors, as performance remains instance-dependent and oracle selection is not deployable. Future work will therefore focus on practical, instance-aware or adaptive initialisation strategies, as well as extending this framework to broader problem classes, circuit families, and hardware-aware settings.

%
%
%
\bibliographystyle{splncs04}
\bibliography{bridgq_refs}
\end{document}